\documentclass[12pt]{article}
\usepackage{bm}
\usepackage{amsmath}
\usepackage{amssymb}
\usepackage{graphicx}
\usepackage{amsmath}
\usepackage{bbm}
\usepackage{comment}

\begin{document}

\baselineskip 15pt

\title{\bf John Stuart Bell: recollections of a great scientist and  a great man }

\author{GianCarlo Ghirardi\footnote{e-mail: ghirardi@ts.infn.it}\\ {\small
Emeritus, Department of  Physics, the University of Trieste,}\\ {\small the Abdus
Salam International Centre for Theoretical Physics, Trieste.}  }

\date{}

\maketitle

\vspace{3cm}
\section{Introduction}

This contribution to the book in honour of J.S. Bell will probably differ from the remaining ones, in particular since only a part of it will be devoted to specific technical arguments. In fact I have considered appropriate to share with the community of physicists interested in the foundational problems of our best theory the repeated interactions I had with him in the last four years of his life, the deep discussions in which we have been involved in particular in connection with the elaboration of collapse theories and their interpretation, the contributions he gave to the development of this approach, both at a formal level, as well as championing it on repeated occasions\footnote{I have also added various pictures taken from letters by John, and/or illustrating important moments of our interactions during his last years.}. In brief, I intend to play here the role of  one of those lucky persons who became acquainted with him personally, who has exchanged important views with him, who has learned a lot from his deep insight and conceptual lucidity, and, last but not least, one whose scientific work has been appreciated by him. 

Moreover, due to the fact that this book intends to celebrate the 50th anniversary of the derivation of the fundamental inequality which bears his name, I will also devote a small part of the text to recall  his clear cut views about the locality issue, views that I believe have not been grasped correctly by a remarkable part of the scientific community.  I will analyze this problem in quite general terms  at the end of the paper. 

\vspace {2cm}
\section{Some of Bell's scientific achievements}
\vspace{0.5cm}
 Bell  got  bachelor's degrees in experimental physics and  in mathematical physics at Queen's University of Belfast  in the years 1948 and 1949, respectively, and a PhD in physics at the University of Birmingham. We all know very well that already at that time he was absolutely unsatisfied with the conceptual structure of quantum mechanics and with the way in which it was taught. This is significantly expressed by the  statement he made during an interview to Jeremy Bernstein: {\it I remember arguing with one of my professors, a Doctor Sloane, about that. I was getting very heated and accusing him, more or less, of dishonesty. He was getting very heated too and said, 'You're going too far'.}  

Then he began his career working at the Atomic Energy Research Establishment at Harwell, Oxfordshire, but he soon joined  the accelerator design group at Malvern. There he met Mary Bell, who he married  in 1954. To summarize the enormous relevance of this event it seems sufficient  to mention that when writing the preface of his collected works on quantum mechanics he stated: {\it I here renew very especially my warm thanks to Mary Bell. When I look through these papers again I see her everywhere.} Subsequently  they moved to CERN, the Centre for European Nuclear Research in Geneva, and  John worked almost exclusively on particle physics and on accelerator design. However, quantum theory was his hobby, perhaps his obsession. And it made him famous. Much more about this in what follows.

Concerning his first scientific activity,  let me stress that  modelling the paths of charged particles through accelerators in these days before electronic computers became available required a rigorous understanding of electromagnetism, and the insight and judgment to make the necessary mathematical simplifications to render the problem tractable on  mechanical calculators, while retaining its essential physical features. Bell's work was masterly.  We cannot avoid mentioning  that  in this period he gave some clear indications concerning the effect of {\it strong focusing} which has played such a relevant role for accelerator science.

 In 1953 we find him, during a year's leave of absence, at Birmingham University  with Rudolf Peierls. In this period he did a work of paramount importance \cite{bell1} producing his version of the CPT theorem, independently of Gerhard L\"{u}ders and Wolfang Pauli, who got all the credit for it\footnote {In M. Veltmann's words: {\it John's article was conceived independently. It is of course very different, and perhaps today more relevant, than the rather formal field theory arguments of L\"{u}ders. In his introduction John acknowledged the paper of L\"{u}ders, and never thought to even suggest that his work had been done ``independently''.}}. Subsequently, both John and Mary  moved to CERN. Here they spent almost all  their careers. 

In 1967 he produced another important piece  \cite{bell2} in elementary particle theory: he pointed out that many successful relations following from current algebra, and in fact current algebra itself, can be seen as a consequence of gauge invariance. However, by far the most important work  by John in the field of elementary particle theory is the 1969 one \cite{bell3} with Jackiv in which  they identified what has become known as the Adler-Bell-Jackiv anomaly in quantum field theory.  This work solved an outstanding problem in the theory of elementary particles  and over the subsequent thirty years the study of such anomalies became important in many areas of particle physics. 

\section{Bell and the foundational problems}
\vspace {1cm}
\subsection{A brief picture}
In spite of the extremely relevant papers mentioned in the previous section, I cannot forget that, as I have stated above,  quantum theory was his hobby, perhaps his obsession. Actually when in 1963-64  he left CERN for SLAC,  he concentrated  his attention almost exclusively to the foundations of such a theory. It is not surprising that  the most relevant theoretical paper which attracted him has been the celebrated EPR paper \cite{epr}, a work which, giving for granted the local nature of physical processes, challenges, in the authors' intentions, the completeness of quantum theory. This conclusion entails that Einstein can be regarded as {\it the most profound advocate of the hidden variables}, as Bell, quoting A. Shimony, made  clear  in ref.\cite{bell4}. So, Bell shifted his attention to hidden variable theories, and, more specifically, to what \cite{bell5} had been for him  {\it a revelation}: Bohmian Mechanics. I would like to stress that this theory was extremely interesting for him for two main reasons: the first was {\it the elimination of indeterminism} but {\it more important, ..., the elimination of any need for a vague division of the world into `system' on the one hand, and `apparatus'  or `observer' on the other.}

Two problems were strictly related to this new perspective. First of all,  J. von Neumann had proved \cite{von} that no deterministic completion of quantum mechanics was possible, in principle. How could this match with the existence and consistency of Bohm's theory? The  essential contribution of John, after his arrival to SLAC, was to show \cite{bell6} that von Neumann's argument was based on a logically not necessary assumption\footnote {The publication of the paper he wrote on this subject, for various reasons, was delayed up to 1966. In connection with this paper two remarks are at order. First, in 1935 Grete Herman had already proved that von Neumann's argument was circular. Her contribution has been completely ignored by the scientific community which made systematic reference to von Neumann's book as ``The Gospel''. Secondly, it is  important to remember that in his paper, while relating his derivation to Gleason's theorem, Bell stresses that in the case of dispersion free states, to avoid a contradiction one must give up the assumption that {\it the measurement of an observable  must yield the same value independently of what other measurements may be made simultaneously}. In brief he has also clarified the unavoidably contextual nature of any deterministic completion of quantum mechanics, a point of remarkable relevance which I have not put into evidence in the main text since I will not make reference to it in what follows. }. The second problem arose from the fact that Bohmian mechanics exhibited a very peculiar feature: it was basically nonlocal. John tried hard to work out a similar theory which was free of nonlocality, but he did not succeed. So he entertained the idea that one might prove that  nonlocal features would characterize any theory whatsoever which reproduces the quantum predictions. With this in mind he conceived and wrote his fundamental paper \cite{bell7} in which he derived his celebrated and revolutionary inequality.

I would like to stress that I consider  this  a result which makes of John one of the greatest physicists of the past century, since he has made crystal clear something that nobody had ever conceived and which implies a radical change in our views about the world around us: Nature is nonlocally causal!

\subsection{The shifty split}

The other problem of quantum theory which, as remarked above, had worried John since his university times, is the one of its resorting to two different dynamical evolution principles: the first one described by the {\it linear and deterministic} Schr\"{o}dinger equation and the second one taking place when measurements are performed and described by the projection postulate of J. von Neumann, which is fundamentally {\it nonlinear and stochastic}. And this is not the only point. The crucial fact is that there is nothing in the theory which marks in any sense the borderline between the range of applicability of the two just mentioned dynamical   principles. This fact worried specifically John and he made systematically reference to it as {\it the shifty split}. His position is wonderfully summarized by his sentence:
\begin{quote}
{\it There is a fundamental ambiguity in quantum mechanics, in that nobody knows exactly what it says about any particular situation, for nobody knows exactly where the boundary between the wavy quantum world and the world of particular events is located.... every time we put that boundary Ð- we must put it somewhere - we are arbitrarily dividing the world into two pieces, using two quite different descriptions ...}
\end{quote}

This  fundamental question, which  had been the subject of a long-lasting and vivid debate between the founders of the theory, is a theme to which Bell returned continuously from the second half of the sixties up to his last days. But this is not the whole story: he also analyzed many of the so called proposed {\it solutions} to this  problem and he proved that almost all of them are characterized by imprecise, vague, verbal assumptions aimed to avoid to face the real contradiction which, in a quantum view, occurs between the {\it waviness of  quanta} and {\it our definite perceptions}.

Actually, it has been just this position which has led him to pay a specific attention, first of all  (i.e. from 1964 on) to Bohmian Mechanics \cite{bohm} and its variants, and, secondly, in the last 5 years of his life, to the collapse model \cite{GRW1} that we (Rimini, Weber and myself) have presented for the first time in 1964 in a very concise form and then discussed in great details \cite{GRW2} in 1965. These facts allow me to pass now to the real core of my contribution: to describe the many interactions we had and how useful they have been for me and my colleagues.

\section{John's interest in our work}

Obviously we  knew very well John and his fundamental contributions, and we considered him by far the greatest scientist in our field (and not only in our field).

Everybody can imagine our surprise and pleasure, after we had sent the preprint of our paper to the CERN library, to receive a letter  from John starting with the following sentence (Fig.1): 
\begin{figure}[h]
\begin{center}
\includegraphics{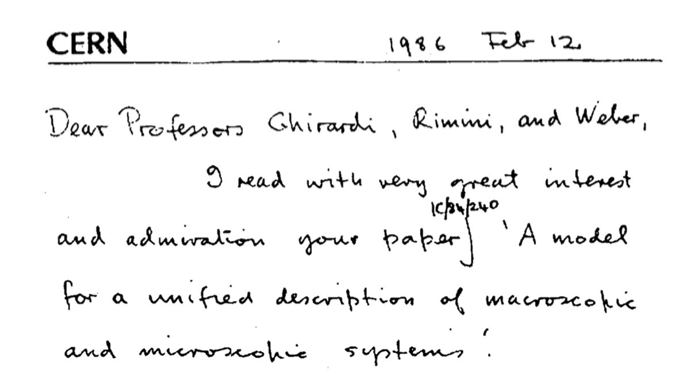}
\caption{The first letter we got from John.} 
\end{center}
\end{figure}

After this significant appreciation,  the scientist  Bell appears with an absolutely appropriate remark (Fig.2):
\begin{figure}[h]
\begin{center}
\includegraphics{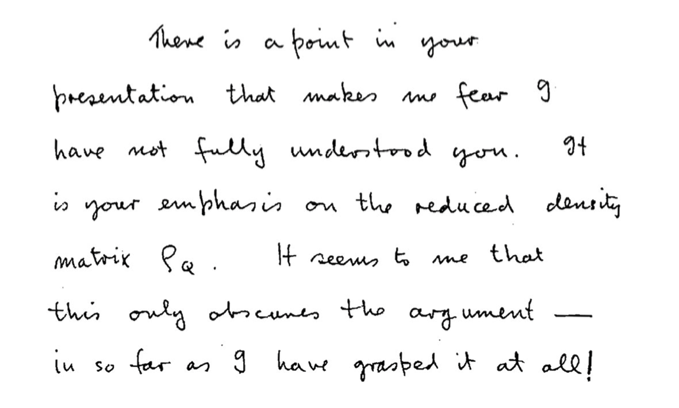}
\caption{A  remark by John.} 
\end{center}
\end{figure}

\begin{figure}[t]
\begin{center}
\includegraphics{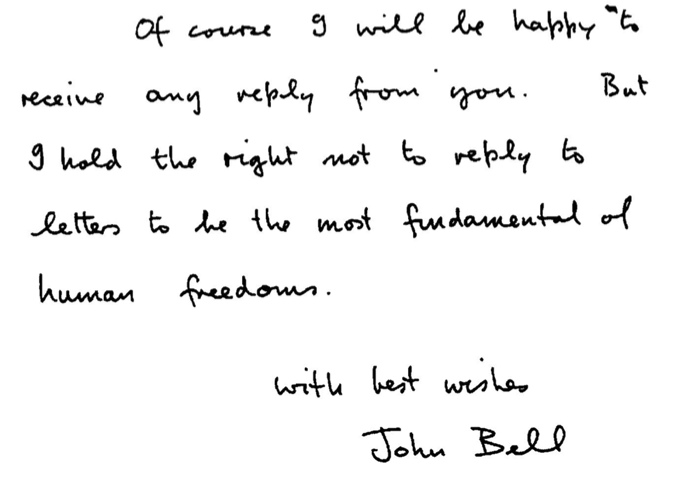}
\caption{The closing sentence of the letter.} 
\end{center}
\end{figure}

The problem is  the one of ensemble versus individual reductions. There is no doubt that our work dealt with individual reductions (and we mentioned this briefly in various parts of the paper), but, 
to perform the complete calculation of the dynamics of a free particle, we resorted to the statistical operator language.  John's letter continues with other general remarks   and   concludes with a sentence (Fig.3) which clearly shows how great and open minded was John as a man: 
\vspace{4cm}

\section{Collapse or GRW models}

Before proceeding, a short summary of what we had done in 1964-5, i.e., to present a proposal  to solve the measurement problem, is at order. Our approach is based on the idea that\cite{bell8}:
\begin{quote}
{\it Schroedinger's equation is not always true.}
\end{quote}

Actually, we suggested to modify  the standard evolution equation  by adding nonlinear and stochastic terms which strive to induce  WPR at the appropriate level leading to states which correspond to definite macroscopic outcomes. The theory, usually referred as the GRW theory, is a rival theory of quantum mechanics and is experimentally testable against it. Its main merit is that it qualifies itself as a precise example of a unified theory governing all natural processes, in full  agreement with quantum predictions for  microscopic processes, and inducing the desired objectification of the properties of macroscopic systems. Let us be  precise about it. 
\begin{itemize}
\item The first problem to be faced is the one of the choice of the so-called preferred basis: if one  wants to objectify some properties, which ones have to be privileged? The natural choice  is the one of choosing the position basis, as suggested by Einstein\cite{einstein}:
\begin{quote}
{\it A macrobody must always have a quasi-sharply defined position in the objective description of reality}
\end{quote}
\item The second problem, and  the more difficult, is to embody in the scheme a triggering mechanism implying that the modifications to the standard theory are absolutely negligible for microsystems while they have  remarkable (and appropriate) effects at the macroscopic level.
\end{itemize}

The  theory is based on the following assumptions: 
\begin{itemize}
\item Let us consider a system of $N$  particles and let us denote as $\psi ({\bf r}_{1},...,{\bf r}_{N})$ the configuration space wavefunction. The particles, besides obeying the standard hamiltonian evolution, are subjected, at random times with a mean frequency $\lambda$, to random and spontaneous localization processes around appropriate positions. If a localization affects the $i$-th particle at point ${\bf x}$, the wavefunction is  multiplied by a Gaussian function $G_{i}({\bf x})=(\frac{\alpha}{\pi})^{3/4}exp[-\frac{\alpha}{2}({\bf r}_{i}-{\bf x})^{2}]$,
\item The probability density of a localization taking place for particle $i$ and at point ${\bf x}$ is given by the norm of the function $G_{i}({\bf x})\psi ({\bf r}_{1},...,{\bf r}_{N})$. This implies that localizations occur with higher probability where, in the standard theory, there is a larger probability of finding the particle,
\item Obviously, after the localization has occurred the wavefunction has to be  normalized again.
\end {itemize}

It is immediate to realize that a localization, when it occurs, suppresses the linear superposition of states in which the same particle is well localized at different positions separated by a distance larger than $1/\sqrt{\alpha}$. 

However the most important feature of the model stays in its trigger mechanism. To understand its basic role let us consider the superposition  of two macroscopically pointer states $|H\rangle$ and $|T\rangle$, corresponding to two macroscopically different locations of the pointer's c.o.m. Taking into account that the pointer is ``almost rigid" and contains a number of the order of Avogadro's number of microscopic constituents one immediately realizes that a localization of any one of them suppresses the other term of the superposition: the pointer, after the localization of one of its constituents, is defintely either Here or There.

With these premises we can  choose the  values of the two constants (which Bell  considered as new constants of nature) of the theory: the mean frequency of the localizations $\lambda$ and their accuracy $1/\sqrt{\alpha}$. These values have been taken (with reference to the processes suffered by nucleons, since it is appropriate to make the frequency $\lambda$ proportional to the mass of the particles) to be:
\begin{equation}
\lambda=10^{-16}sec^{-1},\;\; \frac{1}{\sqrt{\alpha}}= 10^{-5}cm.
\end{equation}
It follows  that a microscopic system suffers a localization, on average, every hundred millions years. This is why the theory agrees to an extremely high level of accuracy with quantum mechanics for microsystems. On the other hand, due to the trigger mechanism, one of the constituents of a macroscopic system, and, corresponingly, the whole system, undergoes a localization every $10^{-7}$ seconds.

Few comments are at order:
\begin{itemize}
\item The theory allows to locate the ambiguous split between micro and macro, reversible and irreversible, quantum and classical. The transitions between the two regimes is governed by the number of particles which are well localized at positions further apart than $10^{-5}$ cm in the two states whose coherence is going to be dynamically suppressed.
\item The theory is testable against quantum mechanics, and various proposals in this sense have been put forward, \cite{{rae}, {rimini},{fu},{penrose2},{adler}, {bassi}}. The tests are  difficult to be performed with the present technology, but  the model clearly identifies  appropriate sets of mesoscopic processes which might reveal the limited validity of the superposition principle.
\item Most of the physics does not depend separately on the two parameters of the theory, but only on their product $\alpha\lambda$ and  a change of few orders of magnitude of its value will already conflict with experimentally  established facts. So, in spite of its appearing ``{\it ad hoc}", if one  chooses to make objective the positions (we mention that one can prove that making objective variables involving the momenta leads to an unviable theory),  not  much arbitrariness remains.
\end {itemize}

An interesting  feature of the theory  deserves a comment. Let us make reference to a discretized version of the model. Suppose we are dealing with many particles and, accordingly, we can disregard the Schr\"{o}dinger evolution of the system because the dominant effect is  the collapse. Suppose  that we divide the universe in elementary cells of volume $10^{-15}cm^{3}$, the volume related to the localization accuracy. Denote as $|n_{1},n_{2},...\rangle$ a state in which there are $n_{i}$ particles in the $i$-th cell and let us consider the superposition of two states $|n_{1},n_{2},...\rangle$ and $|m_{1},m_{2},...\rangle$ which differ in the occupation number of the various cells. It is then quite easy to prove that the rate of suppression of one of the two terms  is governed by the quantity:
\begin{equation}
exp\{-\lambda t \sum_{i}(n_{i}-m_{i})^{2}\},
\end{equation}
\noindent the sum running over all cells of the universe.
\vspace{1cm}

It is interesting to remark that the above equation, being $\lambda=10^{-16}sec^{-1}$, if one is interested in time intervals of the order of the  perceptual times (i.e. about $10^{-2}sec$), implies that the universal dynamics characterizing the theory does not allow the persistence for perceptual times of a superposition of two states which differ for the fact that $10^{18}$ nucleons (a Planck's mass) are differently located in the whole universe.   This remark establishes some interesting connections between the collapse models and the important suggestion by Penrose\cite{penrose} who, to solve the measurement problem by following the quantum gravity line of thought, has repeatedly claimed that it is  the Planck mass which should define the boundary between  the wavy quantum universe  and the one in which the superposition principle fails and, in particular, the world of our definite perceptions emerges.
\vspace{1cm}
\section{John Bell and the GRW model}
\vspace{1cm}

\subsection{Our first personal contacts}

The first time I personally met John has been at the Imperial College (London) at the Centenary celebration  \cite{bell8} of Schr\"{o}dinger. Before this event John wrote to us the  letter shown in Fig.4, in which he  anticipated that he was going to discuss our work.
After having delivered his talk  he  immediately exhibited  his great generosity by telling us: you should have delivered it in place of me!

At that time our proposal had still a big  problem, i.e.,   it did not preserve the (anti)-symmetrization principle for identical constituents.
I remember that during the official dinner of the meeting we (Rimini and myself) discussed with him this problem and he wrote some formulae on a paper napkin. In particular we discussed seriously reductions in which the dynamics strives to make objective the number of particles in an appropriate volume. The collapsing process,  which is accounted for by  the operator  which one should apply to the statevector when a stochastic process occurs, would then have taken the form:
\begin{figure}[t]
\begin{center}
\includegraphics{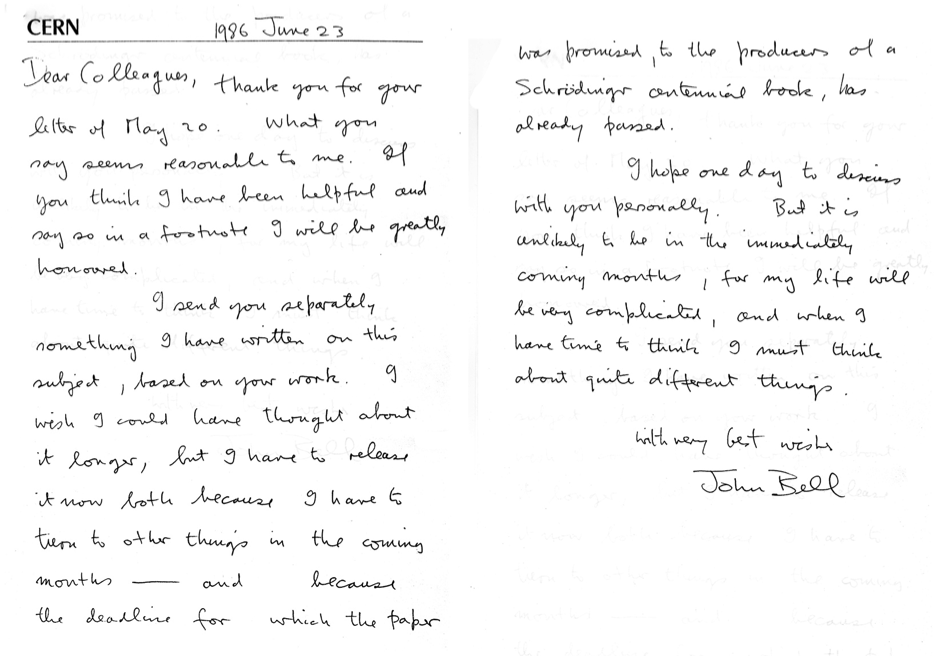}
\caption{John's letter before the Conference in London.} 
\end{center}
\end{figure}

\begin{equation}
|\Psi\rangle\rightarrow e^{-\beta[N({\bf r})-n]^{2}}|\Psi\rangle,\;\;\;N({\bf r})=\int_{V}a^{\dag}({\bf r})a({\bf r})d{\bf r}.
\end{equation}
\noindent Here $a^{\dag}({\bf r})$ and $a({\bf r})$ are the creation and annihilation operator for a particle at point ${\bf r}$ and $V$ is the characteristic localization volume. As usual the statevector has then to be normalized. The reader will easily realize that this process suppresses superpositions of states with different numbers of particles in $V$, and that, being expressed in terms of the creation and annihilation operators, it  automatically respects the symmetry conditions for identical constituents.

The proposal represents a very obvious and simple way to overcome the problem we were facing and from a physical point of view it  leads to results quite similar to those of the original collapse model \cite{GRW2}. However, we  where not fully satisfied with it because it required the introduction of a new parameter in the theory. In fact, besides the frequency of the stochastic processes and the localization volume $V$ it involved the  parameter $\beta$ governing the rate of suppression of states with different number of particles in this volume. We were not very keen to add new phenomenological parameters to the theory. On the contrary John did not feel so uneasy in doing so. In fact, when  I met him few months later in Padova he told me: {\it if you do not write the identical constituents paper I will write it!}

\vspace{1cm}

\subsection{Enters Philip Pearle}

John Bell played also an important role for the subsequent development of the collapse theories. P. Pearle, since a long time, had suggested   \cite{pearle1} that the measurement problem had to be solved by resorting to a stochastic modification of  Scr\"{o}dinger's equation but he had not been able to identify an appropriate preferred basis and a dynamics implying the trigger mechanism.  Just in 1986 Pearle wrote to Bell asking whether he could spend one sabbatical year at CERN interacting with him on foundational problems. Bell replied suggesting him to come to Trieste, and wrote a letter to us (Fig.5) concerning this matter.  
It is interesting to see the reasons that John puts forward for his proposal, reasons which, one one side, make clear how involved he was in other problems and, at the same time, reveal his continuous desire to deal with foundational issues\footnote{It seems appropriate to recall that one time Bell said: {\it I am a Quantum Engineer, but on Sundays I Have Principles.}}.
\begin{figure}[t]
\begin{center}
\includegraphics{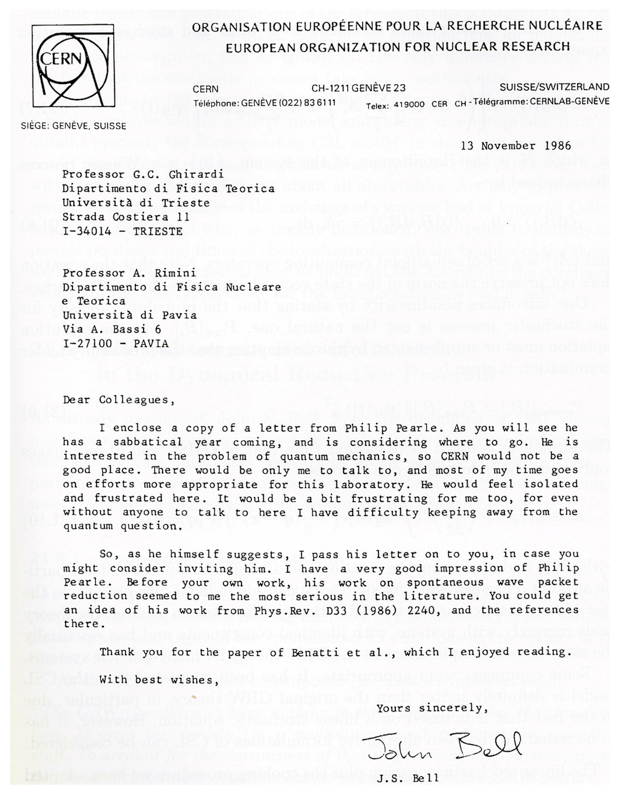}
\caption{John's letter supporting the visit of P. Pearle.} 
\end{center}
\end{figure}
Obviously we reacted immediately and so Philip spent a reasonably long period in Trieste  interacting with me and Renata Grassi. Subsequently he spent few  months in Pavia, invited  by A. Rimini who had got a chair there. This stay of  Pearle has given rise to an extremely useful collaboration between us. But, more important, this contact allowed him to grasp the precise spirit and the technical details of our work, so that, integrating the new ideas in the stochastic evolution equations he was investigating since various years, he produced the elegant version \cite{pearle2} of collapse model which became known as CSL (Continuous Spontaneous Localization).

This proposal, which physically has effects quite similar to  the one worked by us, is formulated in a much more elegant way than GRW and  satisfies the quantum requests for systems with identical constituents, yielding a  solution to our problem without requiring further parameters.

\subsection {The Erice 89 meeting}

The subsequent relevant event in which we met has been the  Conference {\it Sixty-two years of uncertainty} organized by A. Miller in august 1989 at Erice. The readers involved in foundational problems will have no difficulty in identifying some important scientists and philosopher who were present by looking at the photo of the participants (Fig.6) I am attaching below.

 During this meeting the problem of the interpretation (today one would say the ontology) of collapse models has seen an interesting development. To make things clear I will begin mentioning that in his presentation of the GRW theory at Schr\"{o}dinger's Centenary Conference Bell had proposed a very specific interpretation strictly connected with his firm conviction concerning the necessity of making clear what are the ``beables'' of any scientific theory. His proposal has been denoted recently \cite{tumulka} as the {\it flash ontology}. I summarize it by resorting to the precise words he used in {\it Are there Quantum Jumps?}:
 
 \begin{quote}
 The collapse processes {\it  are the mathematical counterparts in the theory to real events at definite places and times in the real world ... (as distinct from the observables of other formulations of quantum mechanics, for which we have no use here). A piece of matter is then a galaxy of such events.}
 \end{quote}

\begin{figure}[h]
\begin{center}
\includegraphics{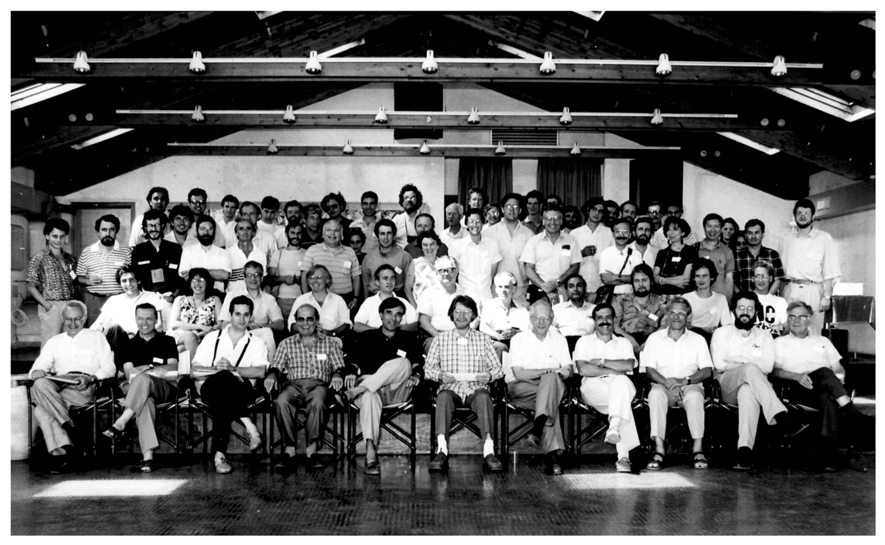}
\caption{Erice's conference in 1989.} 
\end{center}
\end{figure} 

As it is stated clearly, the stochastic processes characterizing the theory are taken as the very basic elements of its ontology. However, in Erice, Bell seemed to have changed his mind by attributing an absolutely privileged role to the wavefunction of a many particle system in the full configuration space:

 \begin{quote}
 {\it The GRW-type theories have nothing in their kinematics but the wavefunction. It gives the density (in a multidimensional configuration space!) of stuff. To account for the narrowness of that stuff in macroscopic dimensions, the linear Schr\"{o}dinger equation has to be modified, in this GRW picture, by a mathematically prescribed spontaneous collapse mechanism. }
 \end{quote}

 I must confess that this is the only point on which I disagreed with John. I am firmly convinced, as many people who are  interested in collapse theories, that they need an interpretation. Limiting all considerations exclusively to the  wavefunction in the 3N-dimensional configuration space does not lead to a clear picture. One needs to connect the mathematical entities with the reality of the world we live in and with our perceptions about it. For this reason we \cite{ggb} proposed  what is presently known as the mass-density ontology (as opposed to the flash ontology), which, at the nonrelativistic level, represents a meaningful way to make sense of  the implications of the theory. What the theory is assumed to be about is the mass density distribution in the real 3-dimensional space at any time, defined as:
 
 \begin{equation}
m({\bf x},t)=\sum_{i}m_{i}\int d{\bf r}_{1}d{\bf r}_{2}...d{\bf r}_{N}|\psi({\bf r}_{1},{\bf r}_{2},...,{\bf r}_{N})|^{2}\delta({\bf r}_{i}-{\bf x}),
\end{equation}

\noindent The mass density interpretation is still at the center of a lively debate involving, among others, philosophers of science\footnote{After having completed our paper \cite{ggb} we sent a copy of it to Bas van Fraassen who answered with the following e-mail: {\it Dear GianCarlo, your message was almost the first I found when I returned here after the holidays, and it makes me very glad. I have talked with my students about your paper and also brought it up in David Albert's seminar (which he was giving here for the fall term). We all agreed that your paper addresses the most important issue about how to relate QM to the macroscopic phenomena in a truly fundamental and new way. .... I will explain below why I see this as part of a consensus with discussions about other interpretations of QM. But there is this difference: that you have given, in your discussion of appropriate and inappropriate topologies, an important and even (to my mind) very convincing rationale for this solution.}}.
However John  stuck, from then on, to the idea that his wave-function ontology is the appropriate one, as I will briefly describe in the next  subsection.

 To conclude this part referring to the Erice's meeting, allow me to present a personal photo taken precisely at Erice. This is  my preferred image among those referring to my professional career since it calls to my mind the exciting moments of that event, and has for me a  deep emotional impact, as any reader would easily understand.

\begin{figure}[h]
\begin{center}
\includegraphics{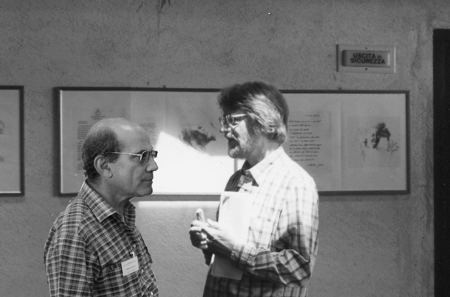}
\caption{John and me} 
\end{center}
\end{figure}

\subsection{More on Bell's ontology}

It goes without saying that I had various other exchanges of view with John about the matter of interpreting the collapse models. Apparently he kept his position, and, in a letter I sent him I gave voice once more to my difficulties with the wave-function ontology and I asked him whether his satisfaction with it was related in a way or another to the fact that one  might extract by it all elements to ground an objective interpretation of things as we perceive them. His clear-cut answer is contained in a letter (Fig.8) he sent to me on october 3, 1989, just one year before his premature death. 

\begin{figure}[h]
\begin{center}
\includegraphics{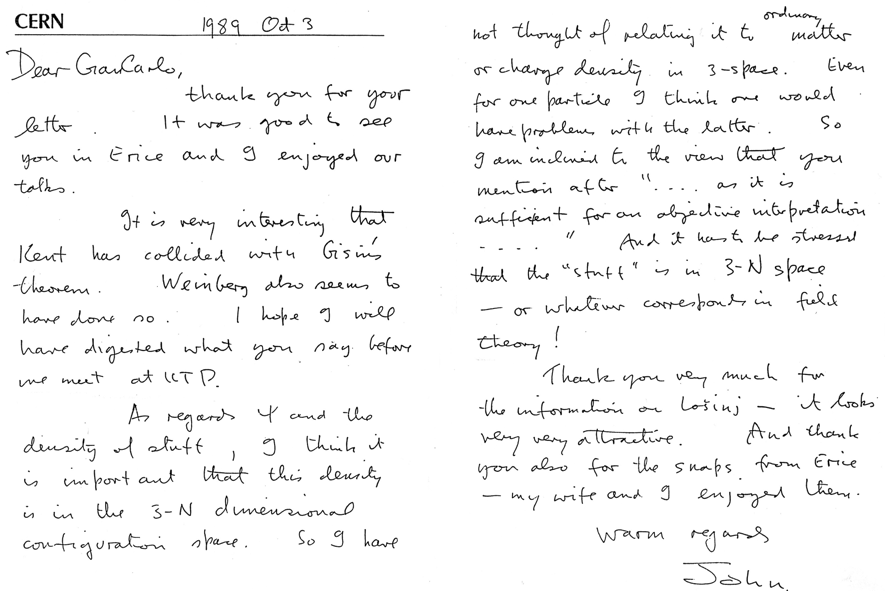}
\caption{Bell's ontology confirmed.} 
\end{center}
\end{figure}

\section{Bell at ICTP}

The last time I met personally John was at the Abdus Salam ICTP, in the fall of 1989, on the occasion of the celebrations for the 25th anniversary of the establishment of this institution. All important speakers were Nobel Prizes, but, fortunately, Abdus Salam was aware of the extreme importance of John's work, and he invited him to deliver a lecture which was chaired by Alain Aspect.
\begin{figure}[h]
\begin{center}
\includegraphics{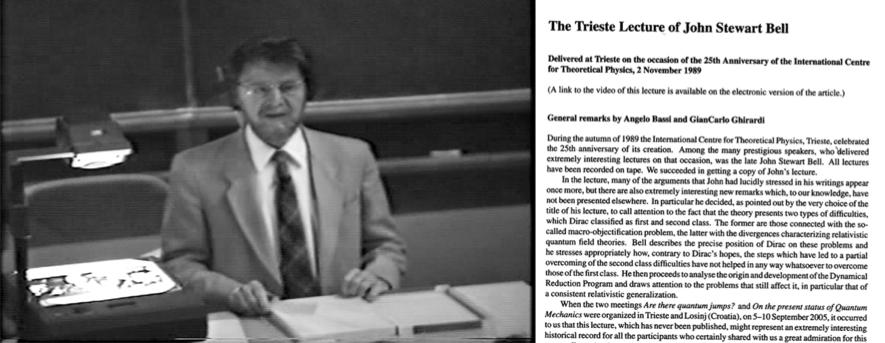}
\caption{John at the beginning of his talk and the introductory remarks to the published version of the same.} 
\end{center}
\end{figure}

I think this was probably the last public general lecture he has delivered,  a wonderful speech by the title: {\it First Class and Second Class Difficulties in Quantum Mechanics}. He went through all fundamental problems of quantum theory, he analyzed the (second class) difficulties connected with the divergences afflicting Quantum Field Theories, the attempts to overcome them, describing the Glashow Salam and Weinberg unification and even commenting on string theories. The second half of his talk was entirely devoted to discuss the first class difficulties (those related to the foundations of quantum mechanics) and, within this context, he  discussed collapse models and the problems connected with their relativistic generalizations.

Just because this  was one of the most stimulating of his talks,  when some of my friends decided to organize a meeting  for my 70th birthday (2005), I worked hard to get access to the video-record of his lecture and  I succeeded in producing a DVD version of it. However, due to the fact that the registration was not ideal (both from the visual and the auditory point of view) and due to our desire of making the talk accessible to all scientists, A. Bassi and myself decided to ``decode'' it and to publish it \cite{bassi2}on the special issue of Journal of Physics A: {\it The Quantum Universe}, collecting papers written by prestigious authors for this occasion.

For the interested reader I am reproducing here an (unfortunately low quality) image of John starting his lecture on november 2, 1989,  as well as the beginning of the presentation we made for his talk. I want  call the attention of the readers on the fact that, even though  various collections of his writings have been published recently,   all of them, unfortunately, missed to include the just mentioned important contribution of his very last years as well as the talk {\it Towards an Exact Quantum Mechanics} he  delivered  on the occasion of the 70¡ birthday of Julian Schwinger,  which has been published by World Scientific in 1989 in: {\it Themes in Contemporary Physics II}, S. Deser and R.J. Finkelstein eds. 

As I have already mentioned, at that time, our attention, as well as the one of all people interested in collapse models, was concentrated on the possibility of getting a relativistic generalization of such models. John had already stressed the fundamental importance of this problem. In fact, few months before, when delivering in Rome a  memorial Bruno Touschek series of lectures he concentrated himself, first of all on the foundational problems of quantum theory, then,  in his second lecture, he discussed Bohmian mechanics and in the third one the collapse theories, as one can deduce by the images I have taken from his transparencies (Fig. 10).

\begin{figure}[h]
\begin{center}
\includegraphics{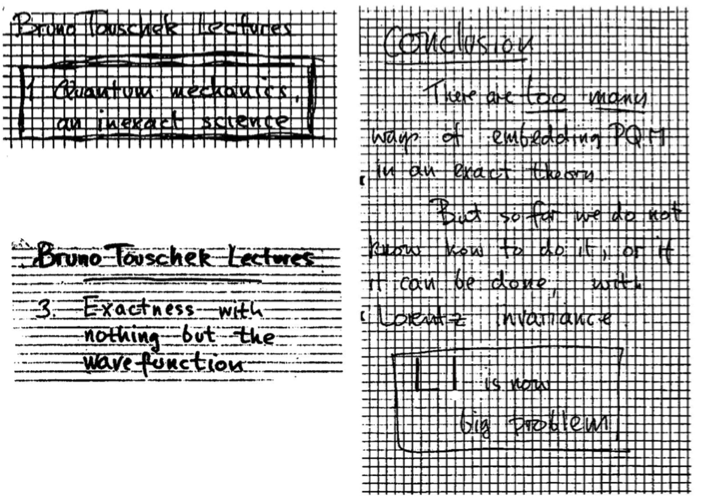}
\caption{Extracts from Bell's transparencies for the B. Touschek lectures.} 
\end{center}
\end{figure}

I consider particularly illuminating that he decided to close his ``short course'' by stressing the paramount importance of working out a consistent relativistic generalization of the theories he had analyzed in the previous lectures, by concluding that  {\bf L}(orentz) {\bf I}(nvariance) is now the big problem. And I cannot avoid mentioning the specific  attention he paid to this problem since the first time he discussed our collapse model  when, after having performed a quite smart analysis of this aspect by resorting to a two-times Scr\"{o}dinger's equation, he concluded (see: {\it Are there quantum jumps?}):
\begin{quote}
{\it For myself, I see the GRW model as a very nice illustration of how, quantum mechanics , to become rational, requires only a change which is very small (on some meaures). And I am particularly struck by the fact that the model is as Lorentz invariant as it could be in the nonrelativistic version. It takes away the grounds of my fear that any exact formulation of quantum mechanics must conflict with fundamental Lorentz invariance}.
\end{quote}
\noindent And in his talk at ICTP, mentioned above,  he stated:

\begin{quote}
{\it There is a whole line of relativistic research here which has been opened up as the Ghirardi-Rimini-Weber jump. And it remains to be seen whether  it will work out well or ill. In any case there is a program where before Ghirardi, Rimini and Weber the fields were rather moribund.}
\end{quote}
\begin{figure}[h]
\begin{center}
\includegraphics{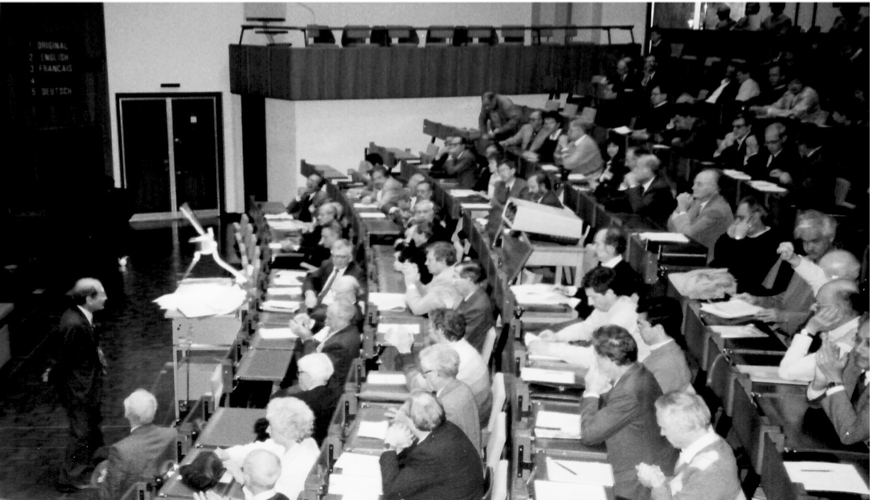}
\caption{Celebration in honor of John at CERN, 1990} 
\end{center}
\end{figure}

Obviously, the problem of getting consistent relativistic generalizations of collapse models (as well as of Bohmian mechanics) has been investigated in many papers in the last years. We do not consider it appropriate to enter here in such a subtle problem. We limit ourselves to call to the attention of the reader that, while all deterministic hidden variable models, and thus Bohmian mechanics , admit relativistic generalizations which require the consideration of a (hidden) preferred reference frame, collapse models admit genuinely Lorentz invariant generalizations. The first one has been introduced by Tumulka \cite{tumulka}, it resorts to the so called {\it flash ontology} and deals with many identical non interacting fermions. Quite recently it has been proved \cite{durr}  that, by resorting to the appropriate way to attribute specific properties to physical systems in a relativistic context with reductions, it is possible to formulate a consistent relativistic model based on the {\it mass density ontology}.

\section{An important celebration}

Just one year after John's death, M. Bell, J. Ellis and D. Amati decided to devote a meeting at CERN to honor  this great scientist. The proceedings have  been published by the Cambridge University Press under the title {\it Quantum Reflections}. The 9 invited speakers have been: R. Penrose, H. Rauch, A. Aspect, G.C. Ghirardi, J.M. Leinaas, A. Shimony, K. Gottfried, N.D. Mermin and R. Jakiv. I really cannot avoid presenting an image,  Fig.11, of such an important event.

John Bell, even after his death, has been celebrated in many other occasions in the following years, and particularly in the present year which marks the 50th anniversary of the derivation of his revolutionary inequality. Just to mention an example, we (D. D\"{u}rr, S. Goldstein, N. Zangh\'{i} and myself) have devoted our 6th annual  meeting on the foundations of quantum mechanics at Sexten precisely to him and to celebrate his inequality.  R. Penrose was happy to take part to this event\footnote{One can look at the registration of all lectures delivered there by going to the web site of the Sexten Centre for Astrophysics (http://www.sexten-cfa.eu/ ) and following the link to the meetings of 2014.}. 

\section{A synthetic comment on Bell's proof of nonlocality}

Here I intend to reconsider very briefly Bell's position concerning nonlocality. The main motivation to do so is that Bell himself has always been fully aware that there have been - and (let me state) there still are - basic misunderstandings concerning the extremely deep conceptual and philosophical implications of his work, even by part of great physicists. Moreover, besides being aware of it, John was also quite upset by this fact.

To start with I will first of all mention an explicit sentence, which appears in {\it Bertlmann's socks and the nature of reality}, in which he has given clear voice to his disappointment for the way in which his work has been interpreted:

\begin{quote}
{\it It is remarkably difficult to get this point across, that determinism is not a presupposition of the analysis ... My first paper on this subject starts with a summary of the EPR argument from locality to deterministic hidden variables. But the commentators have almost universally reported that it begins with deterministic hidden variables.}
\end{quote}

Let me just mention that the term determinism (or equivalently the term realism) is used to assert that the observables of a physical system have definite values (coinciding with one of the eigenvalues of the associated self-adjoint operator) even before the measurement process is performed. On the other hand, and as  well known, the term Hidden Variables  denotes mathematical entities which, either by themselves or in addition to, e.g., the statevector, determine either the precise outcomes of the measurement of any observable (deterministic Hidden Variable Theories), or even only the probabilities of such outcomes (stochastic Hidden Variable Theories).

Given these premises I can rephrase Bell's  argument  in complete generality. Let us consider:

\begin{itemize}
\item A completely general theory such that the maximal - in principle -  specification of the state of a composite system with far apart (quantum-mechanically entangled) constituents determines uniquely the probabilities of all conceivable outcomes of single and correlated measurements, 
\item The state is formally specified by two types of variables $\mu$ and $\lambda$, which are, respectively, accessible and non-accessible. The unaccessible variables are distributed according to an appropriate non-negative distribution (over which averages have to be taken in order to get the quantum probabilities) $\rho(\lambda)$ such that $\int_{\Lambda}\rho(\lambda)d\lambda=1$,
\item The two measurement settings are chosen and the measurement processes are performed and completed in space-like separated regions A* and B*,
\item The specification of the ``initial''  state given by $\mu$ and $\lambda$ refers to a space like surface which does not intersect the common region of the past light cones from A* and B* (so that each region is screened off from the other),
\item The settings can be chosen freely by the experimenters (the free will assumption),
\item The probabilities both of single events as well as of the correlations, coincide with those of Q.M.
\end{itemize}
 
For simplicity let us  make reference to an EPR-Bohm-like situation for a spin singlet with settings $a,b$ and outcomes $A, B$, and let us start with the standard relation for conditional probabilities:
\begin{equation}
P(A,B|a,b;\mu,\lambda)=P(A|a,b;B;\mu,\lambda)\cdot P(B|a,b;\mu,\lambda).
\end{equation}

\noindent Assuming
\vspace{0.5cm}

 {\it Outcome Independence}$\rightarrow P(A|a,b;B,\mu,\lambda)=P(A|a,b;\mu,\lambda)$,
 \vspace{0.5cm}

 \noindent and
\vspace{0.5cm}

{\it Parameter Independence}$\rightarrow P(A|a,b;\mu,\lambda)=P(A|a;\mu,\lambda)$,
\vspace{0.5cm}

\noindent one immediately derives {\bf the fundamental and unique request} by Bell, which we will call Bell's Locality and we will denote as $\{ B-Loc\}$:

\begin{equation}
B-Loc \longleftrightarrow P(A,B|a,b;\mu,\lambda)=P(A|a;\mu,\lambda)\cdot P(B|b;\mu,\lambda).
\end{equation}
\noindent Finally, as already stated,  averaging the probabilities over $\lambda$ one must get the quantum expectation values for the considered state.

The derivation is then straightforward. We know that, in the singlet state, one cannot get the same outcome in both spin measurements if they are performed along the same direction. Now:

\begin{equation}
\int_{\Lambda}\rho(\lambda)d\lambda P(A,A|a,a;\mu,\lambda)=0 \rightarrow P(A,A|a,a;\mu,\lambda)=0, a.e.
\end{equation}
\noindent

\noindent Use of $B-Loc$ implies:
\begin{equation}
[P(A|a,*;\mu,\lambda]\cdot[P(A|*,a;\mu,\lambda]=0.
\end{equation}
\noindent In the above equation we have put an asterisk to indicate that in the region $B^{*}$ ($A^{*}$, respectively) no measurement  (or, equivalently, any measurement whatsoever) is performed. From the above equation we have that one of the two factors of the product must vanish. On the other hand $P(A|a,*;\mu,\lambda)=0, \rightarrow P(-A|a,*;\mu,\lambda)=1$, and, similarly, $P(A|*,a;\mu,\lambda)=0, \rightarrow P(-A|*,a;\mu,\lambda)=1$. Just in the same way, taking into account also the fact that $P(-A,-A|a,a;\mu,\lambda)=0$, one proves that all the individual probabilities  take either the value 1 or the value 0, and, as a consequence of $B-Loc$, the same holds for all the probabilities of correlated events. Concluding, the request that $B-Loc$ holds implies that all probabilities take either the value 1 or zero, i.e.: Determinism (which we will denote as $\{Det\}$).

This is the precise sense of Bell's statement that, in his proof, determinism is {\bf deduced} from the perfect EPR correlations and {\bf not assumed}.

The rest of the story is known to everybody. Defining the quantities $E_{\mu}(a,b)$:
\begin{equation}
E_{\mu}(a,b)=\int_{\Lambda}\rho(\lambda)E(a,b;\mu,\lambda)d\lambda,
\end{equation}
\noindent where
\begin{equation}
E(a,b;\mu,\lambda)=P(A=B|a,b;\mu,\lambda)-P(A\neq B|a,b;\mu,\lambda),
\end{equation}
\noindent one trivially derives Bell's inequality in the Clauser-Horne form:

\begin{equation}
|E_{\mu}(a,b)+E_{\mu}(a,b')+E_{\mu}(a',b)-E_{\mu}(a',b')|\leq 2.
\end{equation}

\noindent Identifying, as requested, $E_{\mu}(a,b)$ with the quantum expectation value $\langle\Psi|\sigma^{(1)}\cdot a\otimes \sigma^{(2)}\cdot b|\Psi\rangle$, one easily shows that, for appropriate directions $(a,a',b,b')$ and for the singlet state, the considered combination violates the bound and can reach the value $2 \sqrt{2}$.

At this point  the logic of the argument should be clear:

\begin{eqnarray}
&&\{Experimental\;Perfect\;Correlations\}\wedge \{B-Loc\}\supset\{Det\}  \nonumber \\
&&\{Det\}\wedge\{B-Loc\}\supset \{Bell's\; Inequality\}  \nonumber \\
&&\{General \;Quantum \;Correlations\}\supset\neg\{Bell's\; Inequality\} \nonumber \\
&&\neg\{Bell's \;Inequality\}\supset \neg\{Det\}\vee \neg\{B-Loc\} \nonumber \\
&&\neg\{Det\}\supset \neg \{Experimental\; Perfect \;Correlations\}\vee\neg\{B-Loc\}.
\end{eqnarray}

\noindent Summarizing: $\{Natural\;Processes\}\supset \neg\{B-Loc\}$, i.e.: {\bf Nature is nonlocally causal}.

\section{Conclusion: a further example of the humaneness of  John }

During a pause of the meeting at ICTP we were walking around the Miramare campus at Trieste and John told me: {\it GianCarlo, you know well how important I consider my interest in foundational problems. However I must state that I think that to devote oneself exclusively to this kind of studies is a luxury. One has also to do something more practical to get paid. This is why at CERN I am so involved in acceleratorÕs physics.}

I replied immediately: {\it John, you are putting me in a very delicate position; in the last 20 years I have worked exclusively in the field of foundations of Quantun Mechanics.}

His answer was immediate and extremely comforting: {\it Oh, no, GianCarlo, you have completely ignored that in these years, as a lecturer, you have trained entire generations of young people in teaching them quantum theory. This fact fully justifies your salary!}

I sincerely believe that this is an appropriate anecdote to conclude this paper which intends to honour the great John Bell.

\end{document}